\documentclass[fleqn,usenatbib]{mnras}

\usepackage{newtxtext,newtxmath}

\usepackage[T1]{fontenc}
\usepackage{ae,aecompl}

\usepackage{graphicx}	
\usepackage{amsmath}	

\usepackage{xcolor}

\DeclareSymbolFont{matha}{OML}{txmi}{m}{it}
\DeclareMathSymbol{\varv}{\mathord}{matha}{118}

\usepackage{tcolorbox}
\newtcolorbox{mybox}{colback=red!5!white,colframe=red!75!black}

\usepackage{makecell}

\title[Detectability of kilonovae]{Detectability of kilonovae in optical surveys: {\em post-mortem} examination of the LVC O3 run follow-up}

\author[A. Sagu\'es Carracedo et al.]{
A. Sagu\'es Carracedo,$^{1}$\thanks{E-mail: ana.sagues-carracedo@fysik.su.se}
M. Bulla,$^{1,2}$
U. Feindt$^{1}$
and A. Goobar$^{1}$
\\
$^{1}$ The Oskar Klein Centre, Department of Physics, Stockholm University, AlbaNova, SE-106 91 Stockholm, Sweden\\fligo
$^{2}$ Nordita, KTH Royal Institute of Technology and Stockholm University, Roslagstullsbacken 23, SE-106 91 Stockholm, Sweden\\
}

\date{Accepted XXX. Received YYY; in original form ZZZ}

\pubyear{2019}

\begin{document}

\label{firstpage}
\pagerange{\pageref{firstpage}--\pageref{lastpage}}
\maketitle

\begin{abstract}
The detection of the binary neutron star (BNS) merger GW170817 and the associated electromagnetic (EM) counterpart, the ``kilonova" (kN) AT2017gfo, opened a new era in multi-messenger astronomy. However, despite many efforts, it has been proven very difficult to find additional kilonovae, even though LIGO/Virgo has reported at least one BNS event during their latest run, O3. The focus of this work is the exploration of the sensitivity of the adopted optical surveys searching for kNe during O3.  We propose ways to optimize the choices of filters and survey depth to boost the detection efficiency for these faint and fast-evolving transients in the future. In particular, we use kN models to explore the dependence on ejecta mass, geometry, viewing angle, wavelength coverage and source distance. We find that the kN detection efficiency has a strong viewing-angle dependency, especially for filters blueward of i-band. This loss of sensitivity can be mitigated by early, deep, observations. Efficient $gri$ counterpart searches for kNe at $\sim200$ Mpc would require reaching a limiting magnitude $m_{\rm lim}=$ 23 mag, to ensure good sensitivity over a wide range of the model phase-space.
We conclude that kN searches during O3 were generally too shallow to detect BNS optical counterparts, even under optimistic assumptions.
\end{abstract}




\section{Introduction}

On August 17th, 2017, during the second run of Advanced LIGO \citep{AdvanceLigo2015} and Advanced Virgo \citep{Acernese2015} collaborations, the first gravitational-wave (GW) signal from the merger of two neutron stars (BNS) was detected, GW170817. Multiple independent detections associated with this event were made throughout the entire electromagnetic (EM) spectrum, including a short gamma-ray burst (\citealt{Abbott2017}; \citealt{Goldstein2017}; \citealt{Savchenko2017}) and the kilonova (kN) AT2017gfo found at optical and IR wavelengths ({\citealt{Coulter2017}};\citealt{SoaresSantos2017}; {\citealt{Valenti17}; \citealt{Arcavi17}};\citealt{Tanvir2017};{\citealt{Lipunov17}}).A record  number of studies were carried out on this single event, and many open questions remain with regards to the rates of neutron mergers, the nucleosynthesis of heavy elements, the mass range and equation of state of neutron stars, the diversity in the kN explosion population and their use as distance yardsticks in cosmology. At only $\sim 40$ Mpc distance, in the nearby galaxy NGC 4993, and close to face-on \citep{Hotokezaka19}, GW170817/AT2017gfo was extremely favourable for detection.

On April 1st, 2019, LIGO/Virgo started the third observing run (O3) on the search of GWs. Since then, astronomers work closely with the LIGO and Virgo collaboration (LVC), who provide search maps with their public alerts used by the astronomy community to carry out target of opportunity (ToO) follow-up programs. One year later, by March 27 2020, the O3 run was prematurely suspended due to the COVID-19 pandemic. The LVC issued alerts for fourteen BNS or NS-BH candidates during O3: one was confirmed as a BNS merger, two were confirmed as a black hole merging with either a low-mass black hole or a high-mass neutron star, 5 are still under review, and the remainder were not astrophysically significant \citep{Abbott2020b}. We explore possible limitations of the adopted optical follow-up strategies, with the aim to guide future efforts by the astronomy community carrying out rapid responses to the LVC alerts.  We give results for a merger distance of 200 Mpc to represent the typical LIGO/Virgo sensitivity during O3. We quantify the kN detection probability for the typical specifications of the follow-up surveys adopted, and suggest potential improvements. Furthermore, we investigate the feasibility of serendipitous kN detection, independently of LVC triggers.

Significant volumes of space are explored with high cadence thanks to the advent of optical sky surveys like PanSTARRS \citep{Kaiser2010}, the All\-Sky Automated Survey for Supernovae \citep[ASAS\-SN,][]{Shappee2014}, the Dark Energy Survey \citep[DES,][]{des}, the Asteroid Terrestrial \-impact Last Alert System  \citep[ATLAS,][]{Tonry2018} or the Zwicky Transient Facility \citep[ZTF,][]{Graham2019} among others. These surveys have also been carrying out regular targeted observations searching for potential EM counterparts of GW alerts in the regions posted by the LVC during O3, whenever the weather and the observable region of the sky made it possible. 

In this work, we combine theoretical kN models with survey simulation tools to explore the detection probability. We focus on optical wavelengths, making use of the ZTF $gri$ filter system, at effective wavelengths 4814, 6422 and 7883 \AA ,  as a representative set of photometric observations. While previous studies used either the observed lightcurve of AT2017gfo or spherically-symmetric kN models to investigate the kN detectability \citep[e.g.][]{Scolnic2017,Setzer19,Rosswog2017,Andreoni2019PASP,Garcia2020}, here we address this problem by generating 3D viewing-angle dependent kN models with the radiative transfer code \textsc{possis} \citep{Bulla2019}. In particular, we choose a {\em fiducial} model that best fits the photometric data of AT2017gfo as our bench mark, also when comparing other models.

In Section \ref{sec:models}, we describe the properties of synthetic kN lightcurves from \textsc{possis}. In Section \ref{sec:surveysimulations}, we present the details of the observing strategies explored, along with the distribution of transients generated and the detection criterion considered. Section \ref{sec:LVCtriggers} focuses on the detection probabilities for the LVC alerts in O3. In Section \ref{sec:synthetic_population}, we describe the population of synthetic kNe selected under various scenarios. Section \ref{sec:rates} considers attainable constraints on kN rates from serendipitous searches, unrelated to LVC alerts. Finally in Section \ref{sec:discussion}, we discuss and summarize the results.

\section{Kilonova models}\label{sec:models}

We consider kN models synthesized using the Monte Carlo radiative transfer code \textsc{possis} \citep{Bulla2015,Bulla2019}. Assuming homologous expansion and analytic functions for the wavelength- and time-dependence of the opacity, \textsc{possis} performs time-dependent radiative transfers simulations for multi-dimensional models. Time-dependent spectral energy distributions (SEDs) are computed for different viewing angles and used to construct multi-band lightcurves.

Compared to the version of \textsc{possis} in \cite{Bulla2019}, here we use an improved version, where the temperature is no longer parameterized and uniform throughout the ejecta, but estimated self-consistently from the mean intensity of the radiation field at each time and in each zone. Furthermore, we adopt thermalization efficiencies $\epsilon_\mathrm{th}$ from \cite{Barnes2016} rather than assuming $\epsilon_\mathrm{th}=0.5$ as in \cite{Bulla2019}. For more information about the code, we refer the reader to \cite{Bulla2019}.

A two-dimensional geometry tailored to BNS mergers is used, comprising two ejecta components as in \cite{Bulla2019}: a lanthanide-rich component distributed within an angle $\pm\phi$ around the merger/equatorial plane, and a lanthanide-free component at higher latitudes (see fig. 1 in \citealt{Bulla2019}). A density profile $\propto \varv^{-3}$, where $\varv$ is the velocity, is adopted for both components between $\varv_\mathrm{min}=0.025$c (instead of $\varv_\mathrm{min}=0$ as in \citealt{Bulla2019}) and $\varv_\mathrm{max}=0.3$c (where c is the speed of light). The three free model parameters explored in this work are:
\begin{itemize}
\item the total ejecta mass, $m_\mathrm{\rm ej}$
\item the half-opening angle, $\phi$
\item the viewing angle, $\Theta$
\end{itemize}

For $m_\mathrm{\rm ej}$, we consider $0.01$ to $0.10$ $M_{\odot}$ with a mass step of $0.01$ $M_{\odot}$, and compute half-opening angles, $\phi$,  from 15 to 75$^\circ$ with steps of 15$^\circ$.
A total number of 50 simulations are generated, each comprising predictions for \mbox{$N_\mathrm{obs}=11$} viewing angles $\Theta$ from pole ($\cos\Theta=1$) to equator ($\cos\Theta=0$). SEDs used in this work are made available at \mbox{\url{https://github.com/mbulla/kilonova_models}}.

\begin{figure*}
    \centering
    \includegraphics[width=\textwidth]{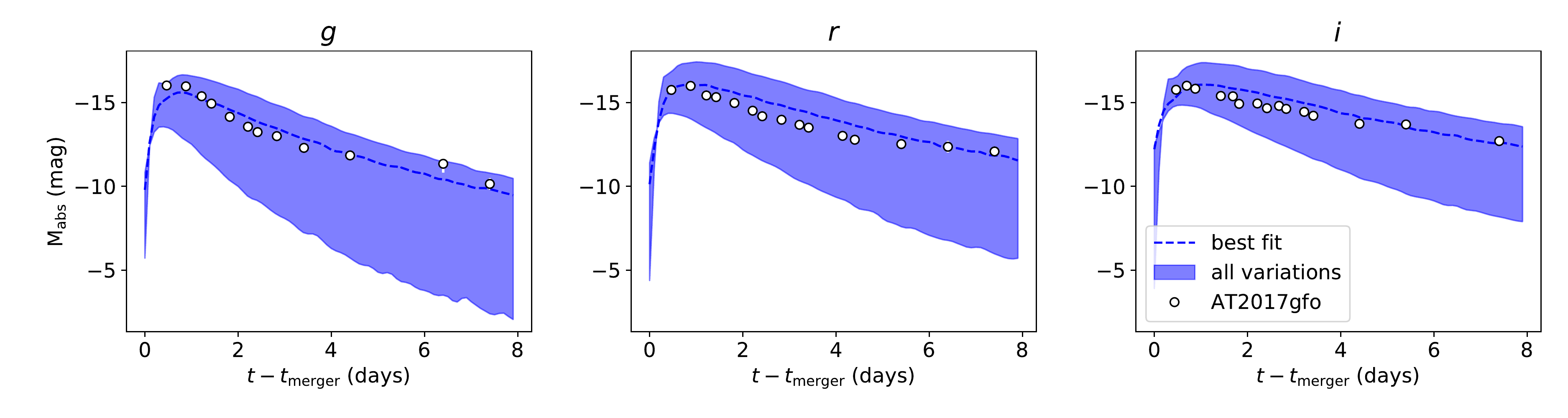}
    \caption{Lightcurve peak magnitudes and shapes allowing for variations in the free parameters of the models: $m_{\rm ej}$ from 0.01 to 0.10\,$\mathrm{M_{\sun}}$, $\phi$ from 15 to 75$^\circ$ and $\Theta$ from 0 to 90$^\circ$ (face-on and edge-on, respectively). The dashed blue line shows the best fit model (fiducial model) that corresponds to $m_{\rm ej}=0.06\,\mathrm{M_{\odot}}$ and $\phi=60^{\circ}$ with  $\Theta=20^{\circ}$. The white dots represent the de-reddened photometric data of AT2017gfo presented in \citet{Smartt17}, \citet{Cowperthwaite2017} and \citet{Andreoni17}.}
    \label{fig:lcs_phot}
\end{figure*}

\subsection{Synthetic lightcurves}\label{sec:lcs}
 The $ugrizyJH$ photometric data of AT2017gfo (\citealt{Andreoni17}; \citealt{Arcavi17}; \citealt{Chornock17}; \citealt{Cowperthwaite2017};
 \citealt{Drout17}; \citealt{Evans17}; \citealt{Kasliwal2017}; \citealt{Pian17}; \citealt{Smartt17}; \citealt{Tanvir2017}; \citealt{Troja17}; \citealt{Utsumi17}; \citealt{Valenti17}) can be reproduced by a {\em fiducial} model with $m_{\rm ej}=0.06\,M_{\odot}$ and $\phi=60^{\circ}$ for viewing angles close to polar, as shown in Fig.~\ref{fig:lcs_phot} for the three filters of interest in this work, $gri$. The figure also shows that kNe are fast-evolving transients with absolute peak magnitudes around -16 mag, i.e., relatively faint compared to many types of supernovae.

\begin{figure}
    \centering
    \includegraphics[width=\columnwidth]{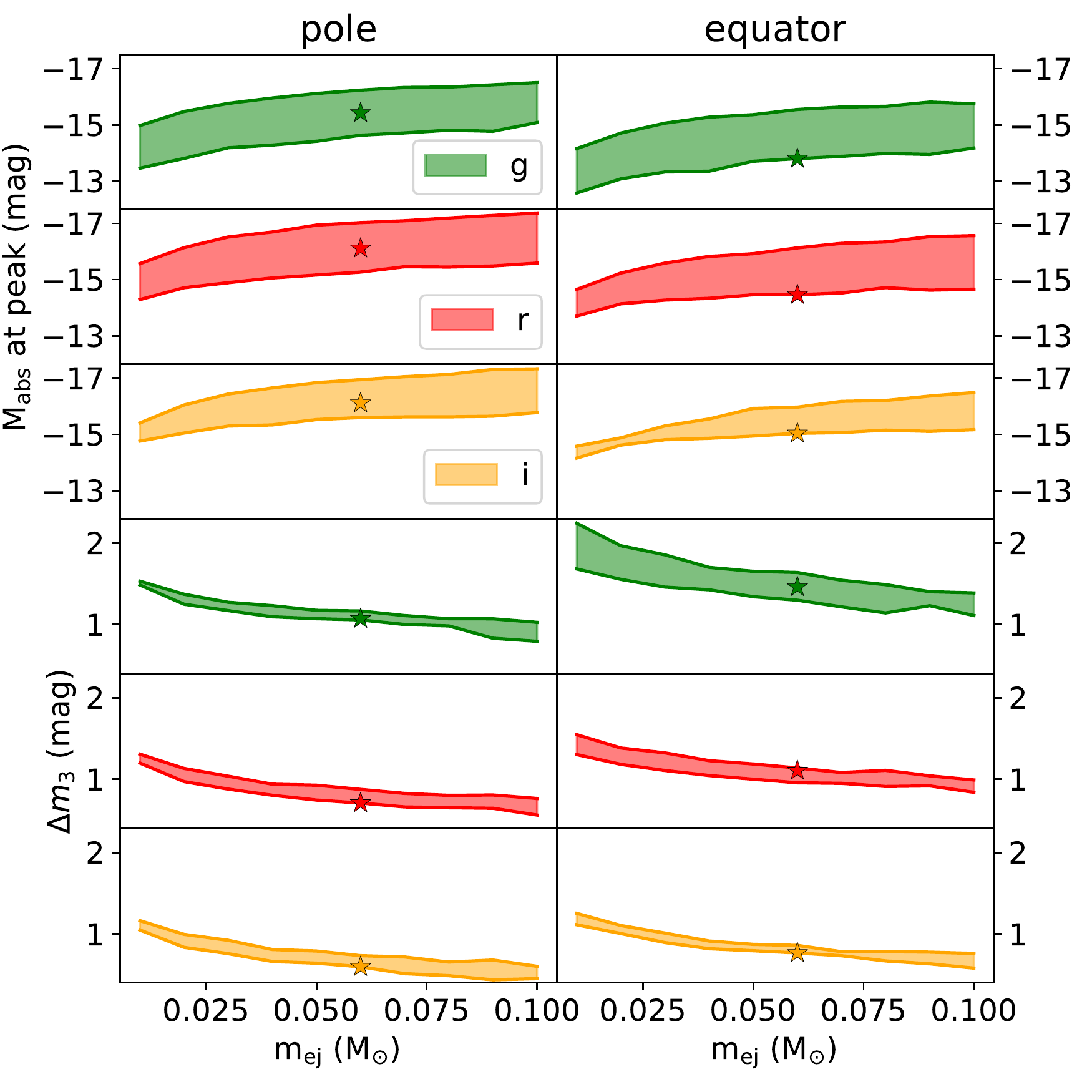}
    \caption{Peak absolute magnitude (three top panels) and decline rate at day 3 after peak (three bottom panels) as a function of the ejecta mass in the models. Synthetic lightcurves at polar (left) and equatorial (right) viewing angles are shown, where the shaded areas represent variations in half-opening angles $\phi$ from 15 to 75$^{\circ}$. The stars indicate the values for the fiducial model.}
    \label{fig:peak_dmag}
\end{figure}

The absolute peak magnitudes as a function of ejecta mass ranges from -12.6 to -17.4 mag, with a decay rate for 3 days after peak ($\Delta m_3$) varying from 0.4 to 2.2 mag, depending on the inclination angle and the broadband filter used, as shown in Fig.~\ref{fig:peak_dmag}. For the fiducial model ($m_{\rm ej}=0.06\,\mathrm{M_{\sun}}$ and $\phi=60^{\circ}$), the optical peak magnitudes vary from -13.8 to \mbox{-16.1 mag},  and $\Delta m_3$ from 0.6 to 1.5 mag for different viewing angles.

\begin{figure}
    \centering
    \includegraphics[width=\columnwidth]{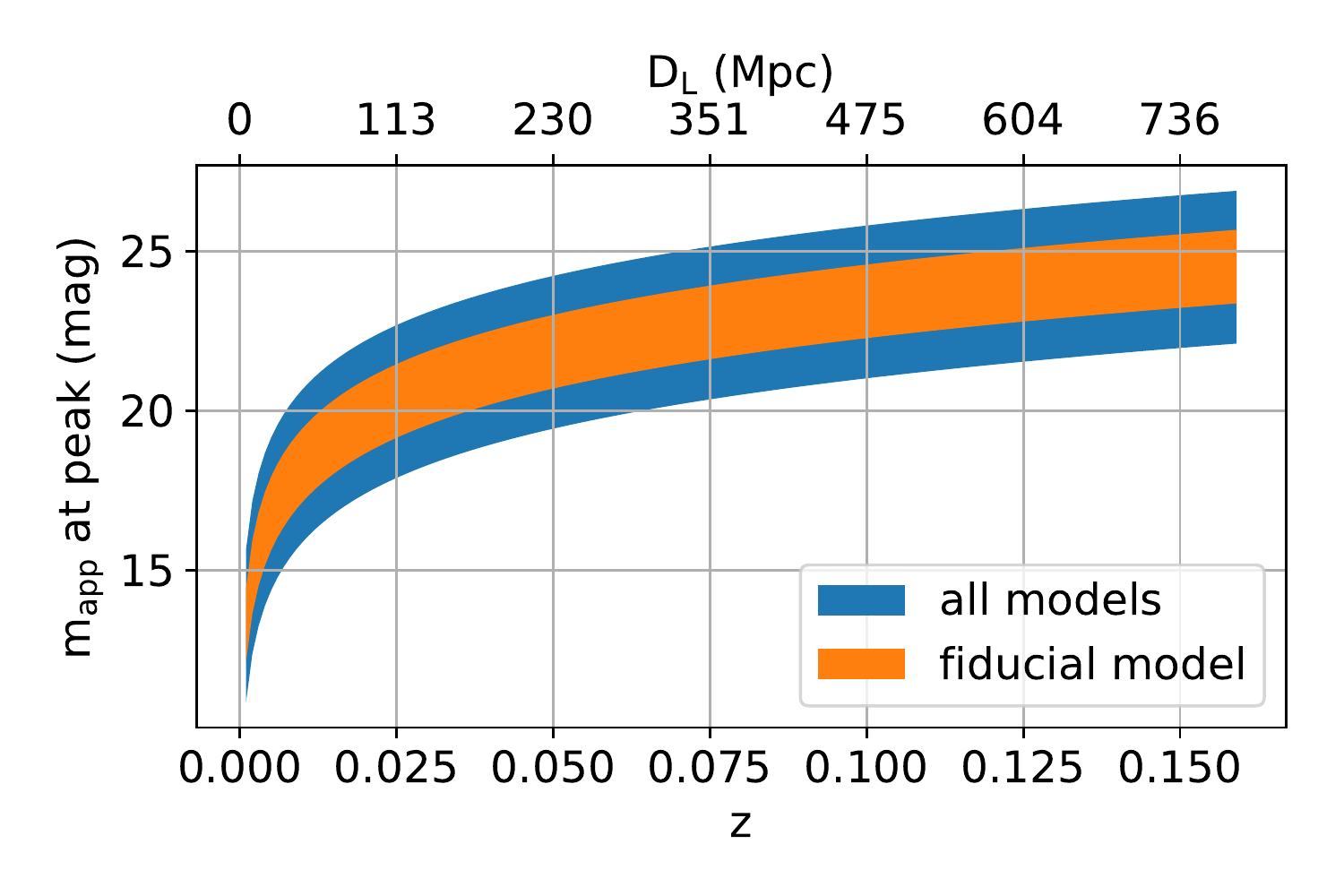}
    \caption{Apparent $r$-band magnitude at peak vs redshift/distance for the entire grid of models. The orange region represents the fiducial model ($m_{\rm ej} = 0.06\,\mathrm{M_{\sun}}
    $ and $\phi=60^\circ$), but for different viewing angles, while the blue region the rest of the models for different $m_{\rm ej}$, $\phi$ and $\Theta$.}
    \label{fig:mag_z}
\end{figure}

Finally, Fig.~\ref{fig:mag_z} shows the observed r-band magnitude as a function of redshift, corresponding to the range of absolute magnitudes at peak, emphasizing the faintness of kNe, and thus the limited redshift range that can be probed with small telescopes. 

Brighter lightcurves result from  larger $m_{\rm ej}$, smaller $\phi$,  and $\Theta$ close to viewing angles towards the pole. The difference in brightness and shape of the synthetic kN lightcurves is less pronounced at redder wavelengths. $\Delta m_3$ is higher for shorter wavelengths and smaller $m_{\rm ej}$, being the difference small when varying $\phi$. 

\section{Survey simulations}\label{sec:surveysimulations}

We perform survey simulations using \texttt{simsurvey}\footnote{\url{https://github.com/ZwickyTransientFacility/simsurvey}} \citep{Feindt2019}, a python based simulation tool for astronomical surveys. \texttt{simsurvey} uses Monte Carlo methods to generate synthetic lightcurves of transients, e.g., supernovae or kNe. It uses a time\-series of SEDs for the transient model, from which the photometry is calculated for a given filter, including cosmological redshift. The input for the simulation is a survey schedule containing the information about cadence, field of view, pointings, filter choice, sky brightness and zero points of each exposure. \texttt{simsurvey} is a tool developed within the ZTF collaboration, and it has been used successfully for forecasting detection rates of supernovae \citep{Feindt2019}.

Besides including the range of kN models with random orientations described above, we also account for dimming by dust in the interstellar medium of the host galaxy.
We follow the procedure used to simulate host galaxy extinction in Type Ia supernovae, among the best studied cases in the nearby Universe \citep{Amanullah2015}. Thus, we  adopt a total-to-selective extinction ratio, R$_v=2$ and a colour excess described by an exponential function \citep[see e.g.][]{Stanishev2018} with scale parameter $\beta=0.11$, as well as Galactic extinction based on the Schlegel, Finkbeiner, Davis (SFD98) reddening maps. In both cases, we use the standard Milky-Way extinction law for the wavelength dependence \citep{Cardelli1989}.

The output of \texttt{simsurvey} is a python dictionary that contains all the parameters of lightcurves for both detected and undetected transients. Examples of the use of \texttt{simsurvey} can be found at \url{https://github.com/ZwickyTransientFacility/simsurvey-examples}. The code used for this work can be found at \url{https://github.com/asaguescar/kne_detectability}.

\subsection{Adopted survey plan}\label{subsec:survey_plan}
We simulate ToO triggers under optimal conditions, i.e., shortly after the merger of two neutron stars. The synthetic observations start between 2.4 to 72 hours after the merger, and continue for seven days, with observations twice per night. The lower limit comes from model coverage and the upper limit is considered to account for whatever reason that might delay the start of the observations up to 3 days, like weather or schedule issues. Although we use ZTF $gri$ filters for the simulations, we expect little dependence on the specifics of these filters, i.e., our findings should be applicable to most optical follow-up campaigns during O3. We simulate lightcurves down to a hypothetical $m_{\rm lim} =$ 25 mag in all filters. Magnitudes are in AB system throughout this work.   

\subsection{Transient generation}\label{subsec:transient_generator}

We generate a fixed number of synthetic kN lightcurves falling in the field of view (FoV) of the instrument. We have picked a ZTF FoV with a mean Milky Way (MW) extinction  of $\rm E (B-V)=0.08$ mag. This extinction value corresponds with the median MW extinction of the whole sky. Transients are placed at distances sampled randomly in space within the FoV (47 sq.~deg.) with a constant volumetric rate up to $z=0.15$. We assume homogeneity to expand the results to the whole sky. 

The kN viewing angle is randomized, i.e., has constant probability for $\cos{\Theta}$. Thus, because of the assumed axial symmetry in the models, viewing angles closer to the merger plane are more probable to happen than those closer to the pole.

\subsection{Detection criterion}\label{sec:detection_criteria}

In dedicated searches of kNe following a GW alert, a basic requirement is non-detection {\em prior} to the merger time, as well as non-detection after sufficient time has elapsed, beyond which the lightcurves are expected to have faded. To have a non-detection we require a {\em prior} visit with no signal passing the threshold. However, in this study we simplify the situation in that we do not simulate data before the merger, nor after the lightcurves fade. Furthermore, we separate the synthetic observations by at least one hour, a common strategy to discard potential false positives from asteroids. We require at least two photometric detections with a SNR $\ge 5$.

\section{LIGO/Virgo alerts in O3}\label{sec:LVCtriggers}

LVC alerts involving neutron stars from LVC in O3, 6 BNS and 8 NS-BH mergers, were followed-up by ground-based optical/NIR telescopes around the globe, following the ability of each telescope to cover the merger location and weather conditions permitting. The searches were performed trying to optimize coverage of the candidate skymap provided by the LVC. Unfortunately, no identification of a kN was made, nor any other EM detection that could be securely associated to the GW signal. After the publication of the second catalog of compact binary coalescences by LIGO and Virgo, (GWTC-2, \citealt{Abbott2020b}), S190425z/GW190425 is the only confident BNS from O3a and the nature of the O3b triggers remain unknown. 
We explore the detection probability for kNe at the distances reported in the LVC alerts, with the main aim to scrutinize the observing strategy.
Since the modelled grid are tailored to BNS mergers (Section~\ref{sec:models}), we will focus on the low-latency BNS classifications.
The efficiency for detecting NS-BH counterparts has to be treated separately, as their lightcurves are predicted to be brighter and longer-lasting compared to those from BNS mergers \citep{Rosswog2017,Barbieri2019,Kawaguchi2020}. 

We seek to understand which is the limiting magnitude required to detect kNe at the distance range suggested by LVC alerts in the recent run. We thus study the detectability at different luminosity distances,  focusing on the results for a merger at 200 Mpc distance, a representative value for the BNS candidates in O3. The only confirmed BNS event during O3a, GW190425, happened at around $159^{+69}_{-72}$ Mpc, consistent with our assumption.

\subsection{Detection efficiencies for fiducial kN model}

\begin{figure*}
    \centering
    \includegraphics[width=.32\textwidth]{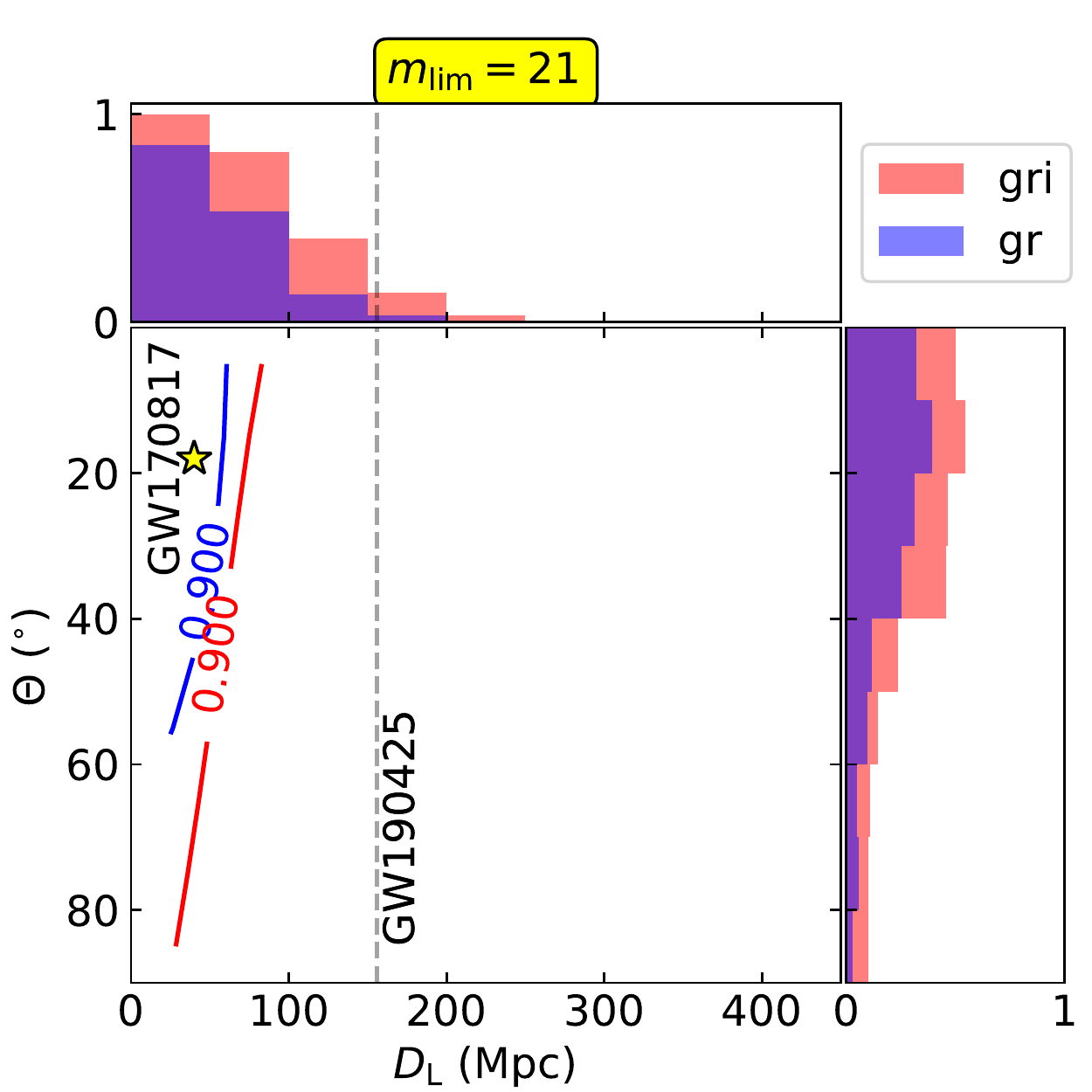}
    \includegraphics[width=.32\textwidth]{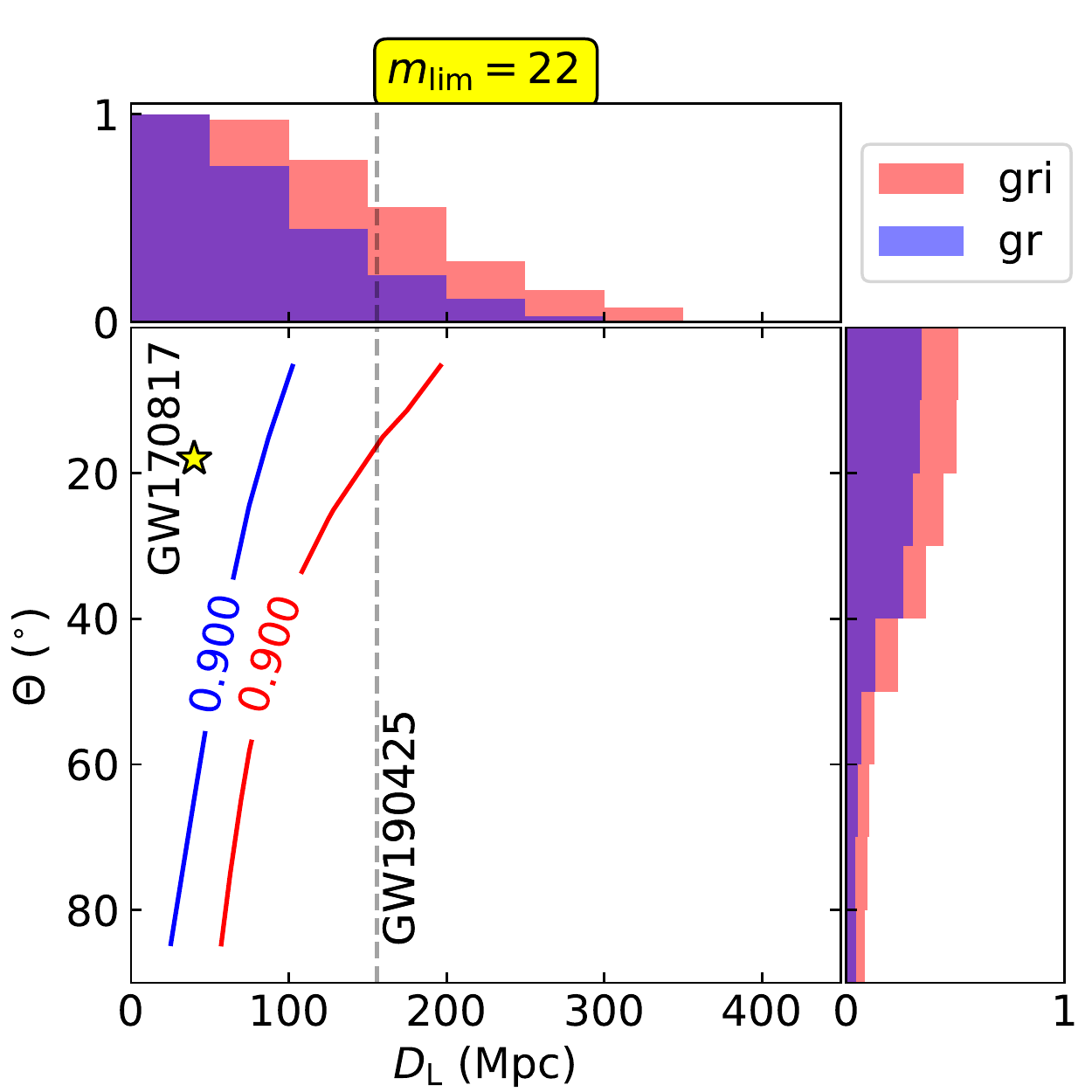}
    \includegraphics[width=.32\textwidth]{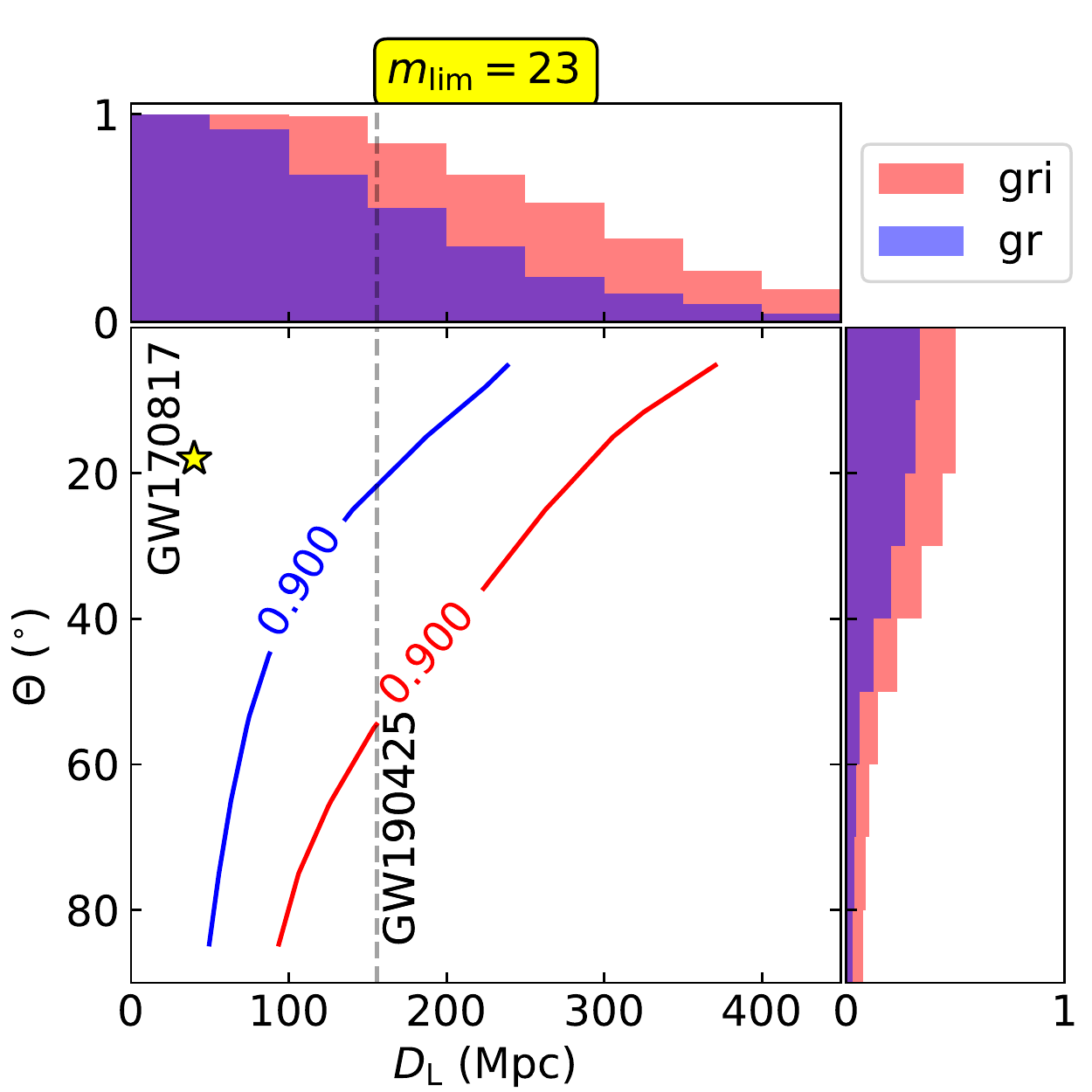}
    
    \includegraphics[width=.32\textwidth]{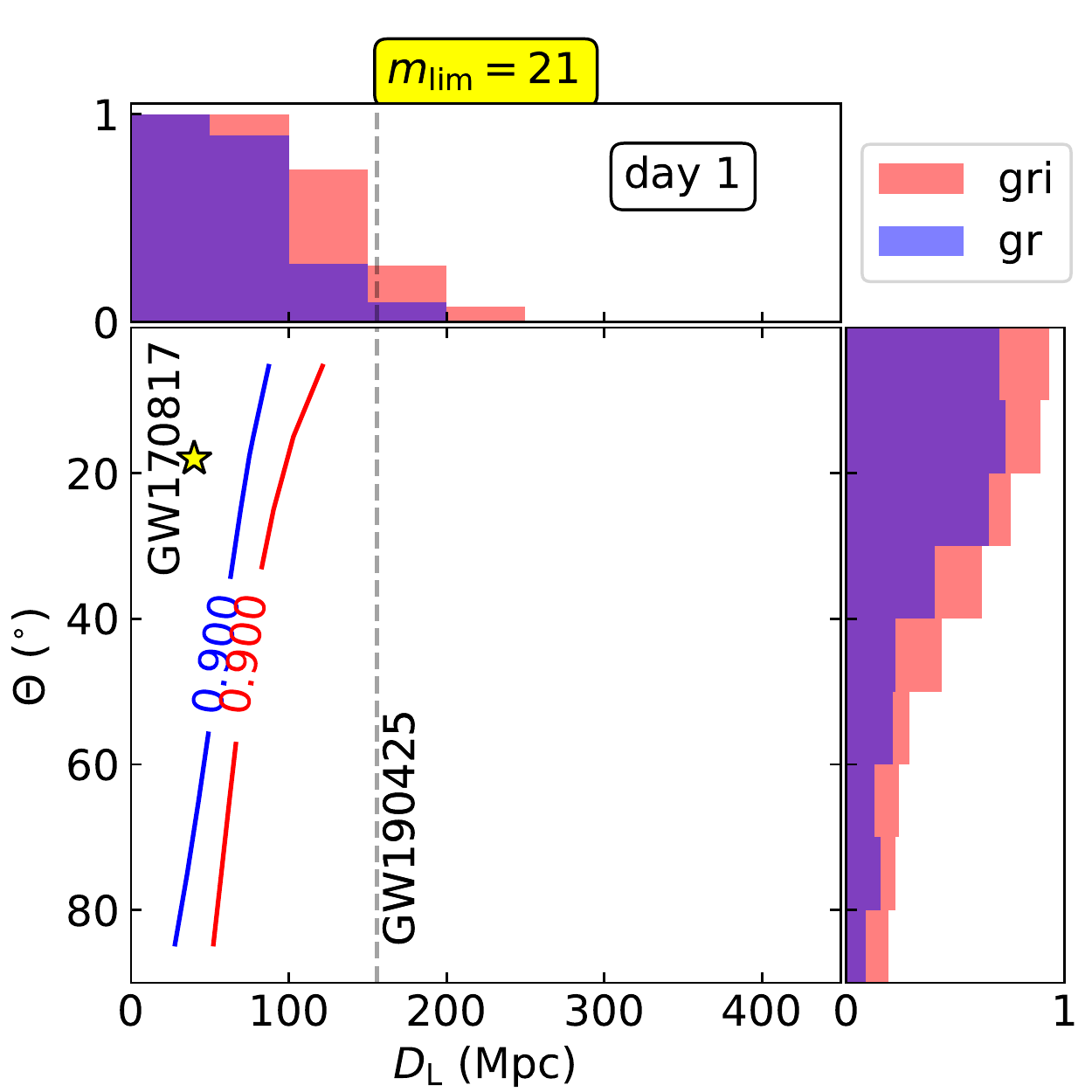}
    \includegraphics[width=.32\textwidth]{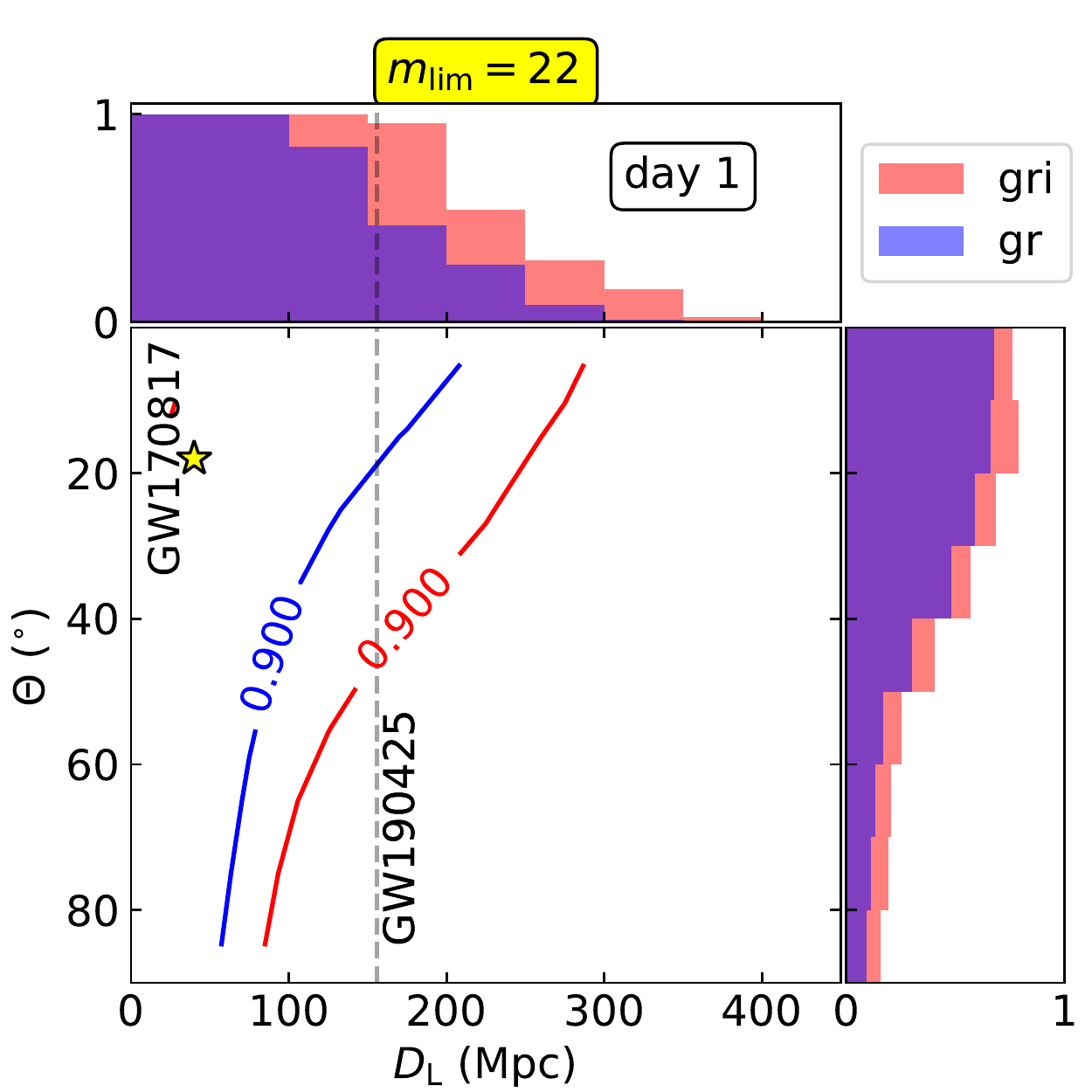}
    \includegraphics[width=.32\textwidth]{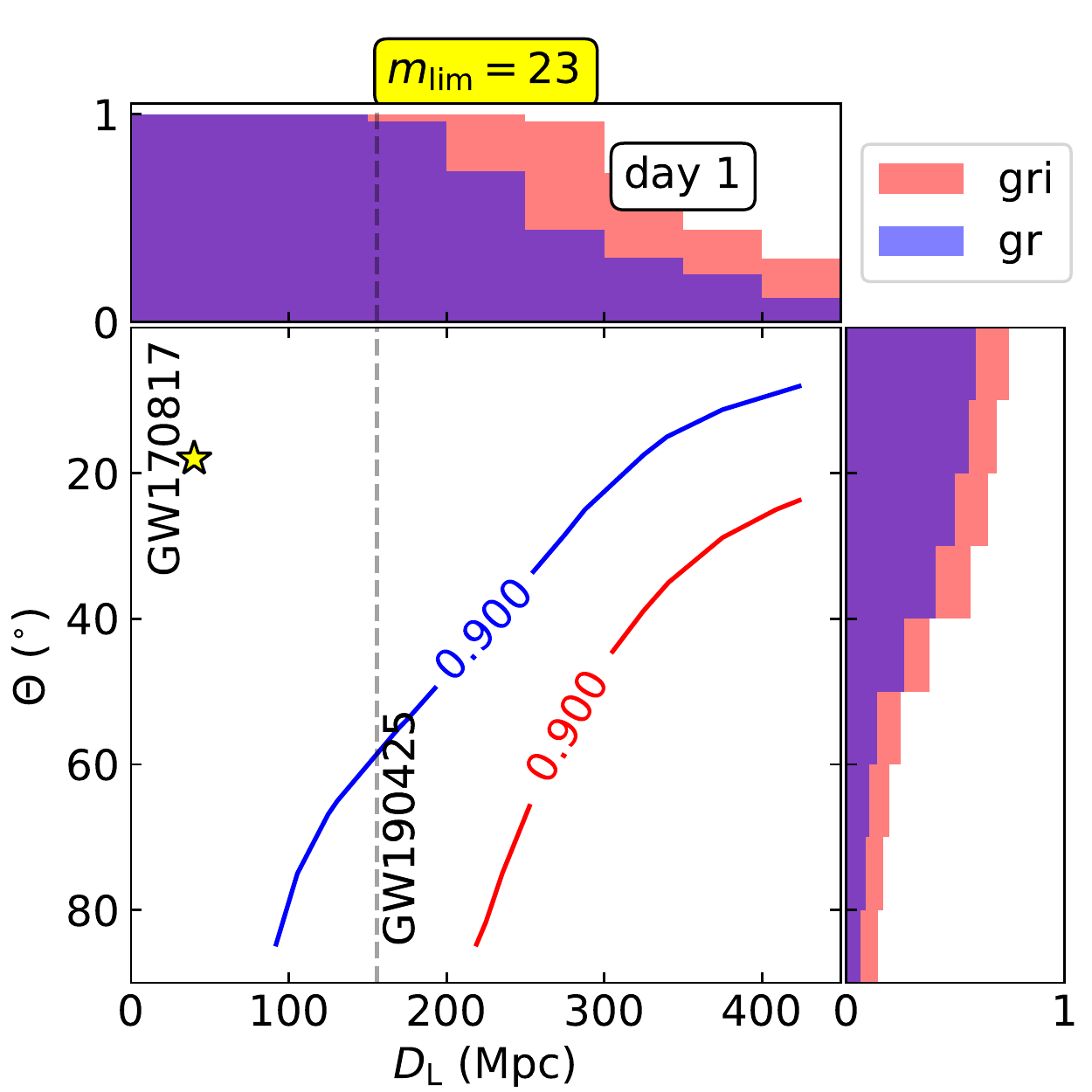}
    
    \caption{Detection probability for the fiducial model (the best match to AT2017gfo). The solid curves show the 2D 90\% probability region, when observing with (red) or without (  {blue}) $i$-band at constant limiting magnitudes 21, 22, and 23 mag (from left to right panel). For each panel, the top and right sub-panels show the projected probability distribution of detecting kN as a function of distance and viewing angle, respectively. Note that a constant depth and observations twice per night are assumed. The yellow star shows GW170817 kN in the adopted parameter space, at a distance of 40 Mpc and viewed from an angle $\Theta = 18^{\circ}$ \citep{Hotokezaka19}. Vertical dashed line indicate the mean estimated distance of the only confirmed BNS event during O3a, GW190425, at around $159^{+69}_{-72}$ Mpc \citep{Abbott2020a} for which the viewing angle is unknown. The three top panels show results for observation starting from 2.4 to 72 hours after merger, while the bottom three panels show the results for observations starting within the first day after merger}.
    \label{fig:hist2deff}
\end{figure*}

The efficiency is calculated from the fraction of the synthetic lightcurve population that fulfill the adopted detection criterion. As expected, the kN detection probability improves for short distances and viewing angles closer to the pole, i.e., where the line of sight points to the lanthanide-free region. Fig.~\ref{fig:hist2deff} shows the detection efficiency for limiting magnitudes $21$, $22$ and $23$ mag, assuming the fiducial model matching AT2017gfo, where the only free parameters are distance and viewing angle. We also indicate the estimated mean distance for the confirmed BNS event GW190425 at $159^{+69}_{-72}$ Mpc \citep{Abbott2020a} as well as the AT2017gfo  kN at the distance of 40 Mpc \citep{Smartt17} and viewing angle $\Theta = 18^{\circ}$ \citep{Hotokezaka19}. The shape of the $90 \%$ probability contours shows the dependency with the viewing angle for the fiducial model, the probability drops significantly towards equatorial viewing angles in which the photons are heavily absorbed by the lanthanides. 
Unlike GW170817, BNS mergers at $\sim200$ Mpc have a low probability of detection at a limiting magnitude of $\lesssim$21 mag,  which was very typical for the O3 follow-up campaigns. For $22$ mag, only favourable viewing angles can be probed, whereas surveys reaching $23$ mag would have had a good chance to detect a counterpart. We also find that redder wavelength coverage, like $i$-band reaching the same limiting magnitude, significantly increases the volume in which a kN is likely observable.

The starting time of the observations affect the detectability, as the earlier the follow-up starts, the higher the chance to identify the rapidly evolving EM counterpart. We investigated the effect of the lag time on the detection probability, finding an improvement of $\sim  10\%$ in the recovery of kilonovae if the follow-up starts already with 24 hs from the merger referred to as "day 1"  and $\sim 5\%$ starting within 48 hs, assuming the source is at 200 Mpc. Fig. \ref{fig:hist2deff} shows the comparison for day 1. 
This means that with low latency observations, kilonovae at a distance of 200 Mpc could be detected by optical surveys with $m_{\rm lim}=22$ mag with high probability for polar viewing angles. Thus the chances of detection are greatly improved compared to the shallower limiting depth adopted by most optical surveys attempting to detected kilonovae O3.

\subsection{Generalized kilonova models}\label{sec:all_grid}

\begin{figure*}
    \centering
    \includegraphics[width=\textwidth]{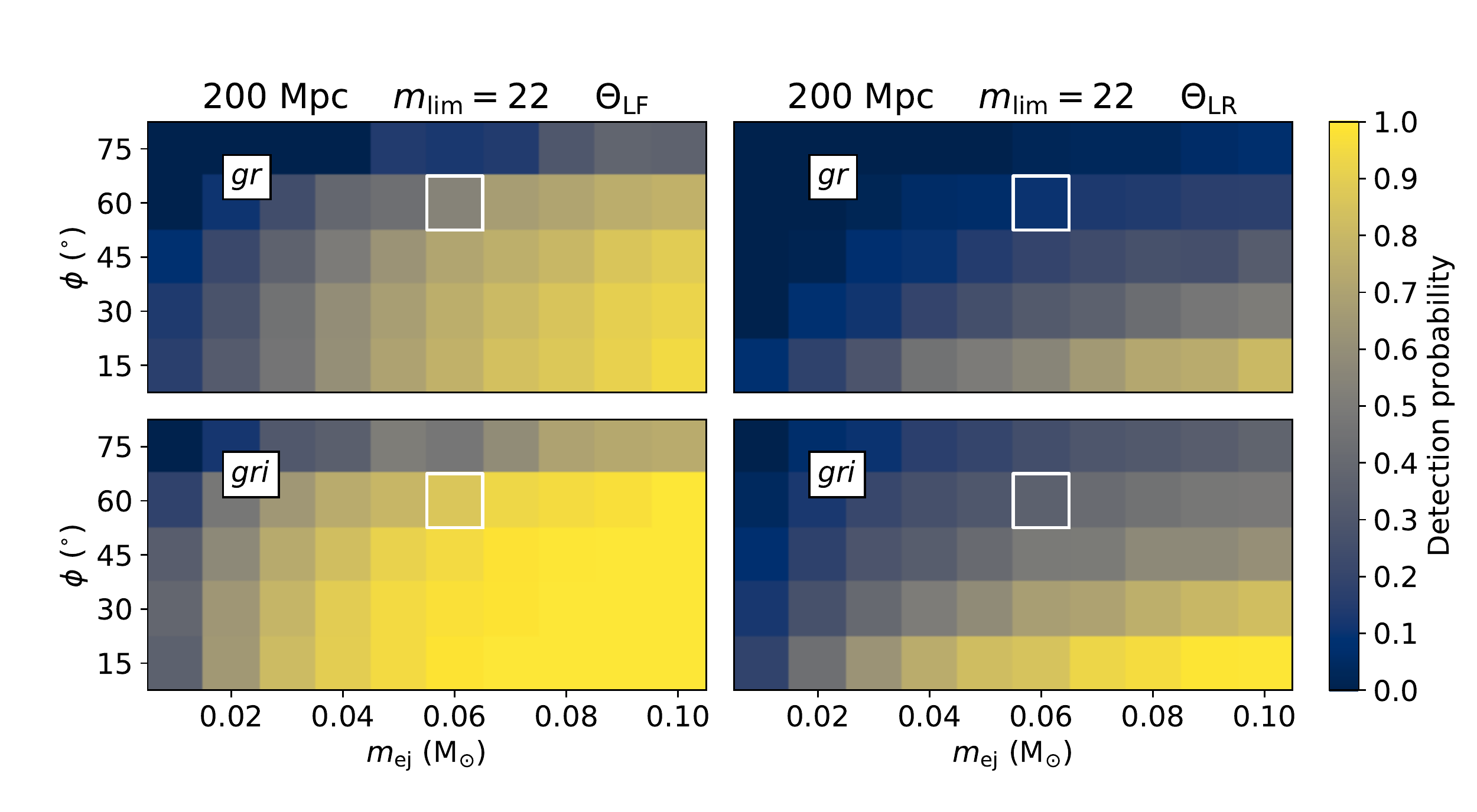}
    \caption{Detection probability for events at $D_{\rm L}= 200$ Mpc, for observations in $gr$ and $gri$ with constant depth of 22 mag and a low Galactic extinction    {$\rm E(B-V)\sim 0.08$}. The left panel assumes viewing angle on the lanthanide-free region, while observations towards the lanthanide-rich region are shown in the right panel. The fiducial model (white rectangle) gives probabilities of 0.87 for $\Theta_{\rm LF}$ with $gri$, 0.54 for $\Theta_{\rm LF}$ with $gr$, 0.36 for $\Theta_{\rm LR}$ with $gri$, and 0.1 for $\Theta_{\rm LR}$ with $gr$ for 22 mag. Only the most favourable viewing angles ($\Theta_{\rm LF}$) give high detection probability for a AT2017gfo-like kN at 200 Mpc.}
    \label{fig:p_22}
\end{figure*}

The probability of kN detection depends on the assumed physical parameters of the model. Fig.~\ref{fig:p_22} indicates the integrated probabilities estimated at 200 Mpc for observations with $gr$ and $gri$, all at an assumed limiting magnitude of 22 mag. Results for 21 and 23 limiting magnitudes can be found in the appendix \ref{app:det_prob}. The probabilities are calculated for two subsets of detections: events with viewing angles pointing to the lanthanide-free component $\Theta_{\rm LF}= \Theta < 90^{\circ}º - \phi$ or to the lanthanide-rich component $\Theta_{\rm LR}=\Theta > 90^{\circ}º - \phi$. Table \ref{tab:det_prob} shows probabilities at 40 Mpc (GW170817), 100 Mpc and 200 Mpc, for the fiducial model and two extremes cases, i.e., models that provide the brightest ($m_{ej}=0.1 M_{\odot}$ and $\phi=15^{\circ}$) and the faintest ($m_{ej}=0.01 M_{\odot}$ and $\phi=75^{\circ}$) kN from the model grid.

The detection probability strongly depends on the physical parameters of the kN and the orientation, i.e.,  large $m_{\rm ej}$, small $\phi$, and $\Theta_{\rm LF}$ are associated with brighter kNe and thus provide higher probability values.

The observations are constraining on the parameter space of the model that provides high detection probability. We aim to be sensitive to a wide range of parameters with our observations. The inclination of the merger is unknown when the ToO is triggered. Viewing angles close to face-on provide the brighter lightcurves, but they are less likely to happen (see section \ref{subsec:transient_generator}). Therefore, it is preferable to plan for less optimal circumstances. Reaching 23 mag depth in $gri$ is needed to constraint most of the parameter space at 200 Mpc.

\begin{table*}
	\centering
	\caption{Detection probability for events with orientation such that the line of sight points to the lanthanide-rich component (LR) and the lanthanide-free component (LF), observing with $gr$ or $gri$, assuming constant depth of 21, 22 and 23 mag at 40, 100 and 200 Mpc. Probabilities are shown for the fiducial, brightest and faintest models from left to right. The chosen distances represent the location of GW170817 ($\sim40\,$Mpc), a representative luminosity distance of O3 LVC events ($\sim200\,$Mpc) and an intermediate distance ($100\,$Mpc).   }
	\label{tab:det_prob}
	\begin{tabular}{cc|cccc|cccc|cccc}
	    \hline
	    \multicolumn{2}{c|}{} & \multicolumn{4}{c|}{fiducial} &  \multicolumn{4}{c|}{brightest} &  \multicolumn{4}{c}{faintest}  \\
	    
		Mpc &lim &   LF$_{gr}$   & LR$_{gr}$  &  LF$_{gri}$  &  LR$_{gri}$&   LF$_{gr}$   & LR$_{gr}$  &  LF$_{gri}$  &  LR$_{gri}$&   LF$_{gr}$   & LR$_{gr}$  &  LF$_{gri}$  &  LR$_{gri}$\\
		\hline
40 & 21 & 1.00 & 0.73 & 1.00 & 0.95 & 1.00 & 1.00 & 1.00 & 1.00 & 0.74 & 0.29 & 0.83 & 0.57  \\
40 & 22 & 1.00 & 0.90 & 1.00 & 1.00 & 1.00 & 1.00 & 1.00 & 1.00 & 0.84 & 0.50 & 0.90 & 0.80  \\
40 & 23 & 1.00 & 0.99 & 1.00 & 1.00 & 1.00 & 1.00 & 1.00 & 1.00 & 0.87 & 0.71 & 0.98 & 0.94  \\
\hline
100 & 21 & 0.79 & 0.27 & 0.96 & 0.59 & 0.98 & 0.90 & 1.00 & 1.00 & 0.17 & 0.00 & 0.30 & 0.11  \\
100 & 22 & 0.97 & 0.56 & 1.00 & 0.89 & 1.00 & 1.00 & 1.00 & 1.00 & 0.33 & 0.06 & 0.51 & 0.40  \\
100 & 23 & 1.00 & 0.81 & 1.00 & 1.00 & 1.00 & 1.00 & 1.00 & 1.00 & 0.46 & 0.31 & 0.73 & 0.68  \\
\hline
200 & 21 & 0.12 & 0.00 & 0.36 & 0.04 & 0.61 & 0.37 & 0.91 & 0.75 & 0.00 & 0.00 & 0.00 & 0.00  \\
200 & 22 & 0.54 & 0.10 & 0.87 & 0.36 & 0.95 & 0.81 & 1.00 & 1.00 & 0.00 & 0.00 & 0.00 & 0.00  \\
200 & 23 & 0.92 & 0.38 & 1.00 & 0.76 & 1.00 & 0.92 & 1.00 & 1.00 & 0.06 & 0.03 & 0.31 & 0.24  \\
\hline
	\end{tabular}
\end{table*}

\section{Synthetic kilonova population detected in simulations}\label{sec:synthetic_population}

Next, we explore the model properties for the population of kNe that pass the detection criterion for a given magnitude limit. We group the viewing angles of the detected sample in two sets: one with orientation such that the line of sight points to the lanthanide-free component ($\Theta_{\rm LF}$) and the other pointing to the lanthanide-rich component ($\Theta_{\rm LR}$), defined in Section \ref{sec:all_grid}. We also separate the detections for synthetic observations into those that include $i$-band and those that do not.
To better assess the selected samples under different considerations, we scale the number of detections to a BNS rate of 1000 Gpc$^{-1}$yr$^{-1}$ \citep{Scolnic2017} for 20\,000 sq. degrees. LVC recently updated their estimate of the BNS merger rate to $320^{+490}_{-240}$ Gpc$^{-1}$yr$^{-1}$ in \citealt{Abbott2020b}. Nevertheless, we have adopted the somewhat larger rate of 1000 Gpc$^{-1}$yr$^{-1}$, which can be easily re-scaled to different rate assumptions and future measurements. 
The summary of our findings is shown in Table \ref{tab:output_sim}.

\begin{table*}
  \renewcommand{\arraystretch}{1.5}
	\centering
	\caption{Summary of simulation statistics. $n_{{\rm KNe}, gr}$ and $n_{{\rm KNe}, gri}$ correspond to the number of detections scaled to a given BNS rate of 1000 Gpc$^{-3}$yr$^{-1}$ \citep{Scolnic2017} for 20\,000 sq. degrees.  $f_{{\rm LF}, gr}$ and $f_{{\rm LF}, gri}$ correspond to the fraction of detections with viewing angles towards the lanthanide free component ($f_{\rm LF}$) observing with $gr$ and $gri$. $\bar{\Theta}_{gr}$ and $\bar{\Theta}_{gri}$ are the mean viewing angle in degrees, and $\bar{D}_{{\rm L}, gr}$ and $\bar{D}_{{\rm L}, gri}$, the mean luminosity distance in Mpc of the detected kNe. The parentheses show the parameters for the extreme models from the model grid (see Sec.\ref{sec:lcs}), the left value corresponding to the faintest model and the right one to the brightest.} 
	\label{tab:output_sim}
	\begin{tabular}{ccccccccc}
		\hline
		$m_{\rm lim}$ & $n_{{\rm KNe}, gr}$ & $n_{{\rm KNe}, gri}$ &$f_{{\rm LF}, gr}$ & $f_{{\rm LF}, gri}$ & $\bar{\Theta}_{gr}$ & $\bar{\Theta}_{gri}$ & $\bar{D}_{{\rm L}, gr}$ & $\bar{D}_{{\rm L}, gri}$ \\
		\hline
20.5 & 1 (0,4)  & 3 (0,8) & 0.42 (0.00,0.83) & 0.30 (0.00,0.83)    & 38 (54,50) & 44 (60,51) & 78 (39,154) & 96 (51,188) \\
 
21.0 & 2 (0,8) & 5 (0,15) & 0.40 (0.00,0.83) & 0.33 (0.00,0.83) & 39 (47,51) & 43 (55,51) & 100 (41,196) & 125 (67,237) \\
 
22.0 & 8 (0,26) & 20 (1,55) & 0.37 (0.04,0.84) & 0.33 (0.02,0.84) & 40 (55,51) & 42 (59,51) & 156 (75,299) & 196 (100,370) \\
 
23.0 & 28 (3,86) & 71 (10,141) & 0.39 (0.04,0.83) & 0.33 (0.04,0.79) & 39 (54,52) & 43 (55,55) & 245 (111,442) & 302 (159,498) \\
 
24.0 & 96 (9,461) & 213 (37,540) & 0.37 (0.06,0.78) & 0.28 (0.05,0.75) & 40 (53,55) & 45 (54,57) & 369 (174,509) & 432 (253,528) \\
 
25.0 & 242 (38,452) & 411 (134,458) & 0.26 (0.07,0.75) & 0.18 (0.05,0.74) & 46 (52,57) & 53 (54,57) & 457 (288,530) & 501 (400,531) \\
		\hline
	\end{tabular}
\end{table*}

Given the red nature of kNe, including $i$-band to the same depth as $g$ and $r$ in the survey leads to more than 50\% percent improvement on the number of detections. kN lightcurves show a slower decay rate in redder bands (smaller $\Delta m_3$, see Fig. \ref{fig:peak_dmag}), they are less affected by the flux suppression of the lanthanide-rich component and are less affected by dust extinction.

The detections increase with the limiting magnitude as more volume becomes available, \mbox{$V \propto 10^{\,0.6\cdot m_\mathrm{\rm lim}}$}.
In our simulations, we find that $g$-band observations, although useful for background rejection and characterization of the kN properties, do not add to the detection efficiency. All the kNe detected with SNR $\ge 5$ do also pass the detection criterion in $r$ and $i$-band.

\subsection{Selection effects}\label{sec:selection_effects}

\begin{figure*}
    \centering
    \includegraphics[width=\textwidth]{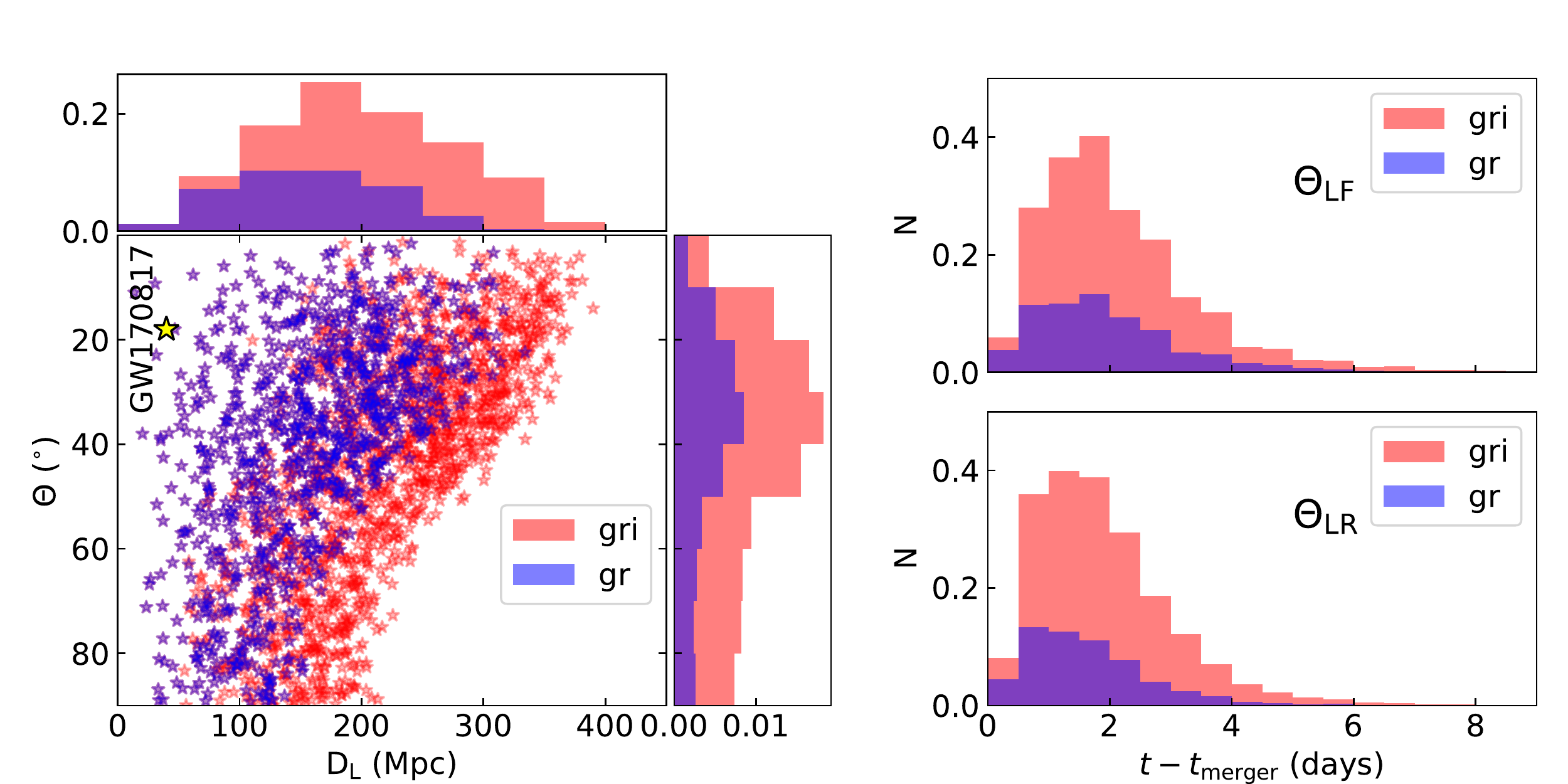}
    \caption{kN detections considering a population with the physical parameters of the fiducial model $m_{\rm ej}=0.06\,\mathrm{M_{\sun}}$ and $\phi=60^{\circ}$. The main left panel shows the detections, distributed in distance and viewing angle, while the smaller panels on the top and on the side show the normalized 1D distributions in distance and viewing angle, respectively. The yellow star shows GW170817 kN in the adopted parameter space, at a distance of 40 Mpc and viewed from an angle $\Theta = 18^{\circ}$ \citep{Hotokezaka19}. 
    The two panels on the right show the distribution of post-merger epoch of the synthetic data points that pass the threshold for the adopted detection criterion (SNR $\ge 5$). Observations with $gr$ (blue) and $gri$ (red) for viewing angles in the lanthanide-rich (bottom) and the lanthanide-free (top) components are shown. The normalized distribution of post-merger epoch does not change significantly for different limiting magnitudes, while the distribution on detection in distance and viewing-angle space will be shifted. We show results for $m_{\rm lim}=22$ mag as a representative example.
    }
    \label{fig:hist2d}
\end{figure*}

The distribution of detections in viewing angle and distance is shown in Fig. \ref{fig:hist2d}.
The detected population is affected by Malmquist bias \citep{Malmquist1922}, i.e., given a magnitude limit, only the brightest objects in the population are detected. For the synthetic population, as volume increases with distance, most kNe are generated at farther distances. Furthermore,  equatorial viewing angles are more common (see Section \ref{subsec:transient_generator}). Thus, the brightest events correspond to lower distances and polar viewing angles. The combination of the two effects gives a median viewing angle of about 40$^{\circ}$ for the population of selected kNe, assuming the fiducial model.

\begin{figure*}
    \centering
    \includegraphics[width=\textwidth]{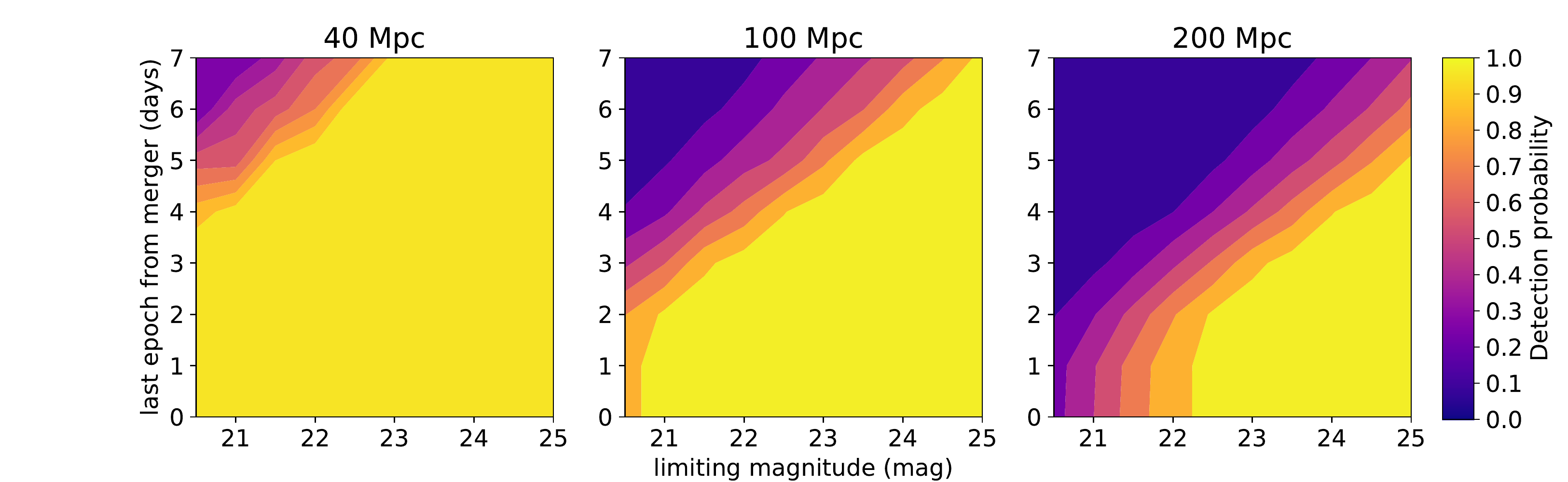}
    \caption{Probability map of detecting kilonovae at 40, 100 or 200 Mpc with constant depth from 20.5 to 25 mag (horizontal axis), for a minimum last post merger epoch required (vertical axis). This plot shows results for the fiducial model with observations including $gri$ and viewing angle uniformly distributed in $\cos{\Theta}$ as described in Section \ref{subsec:transient_generator}. }
    \label{fig:last5p}
\end{figure*}

When searching for transients in the survey alerts, astronomers look for at least two data points with a sufficient signal to noise and time separation to consider a reliable discovery. The chances to observe a lightcurve that can be identified as a kN increase when the transient is observable for a more extended period. The duration of the observable lightcurve depends on the properties of the kN and vary with distance.
We investigate how the observable time window for the kNe lightcurve affects the detection efficiency. For that, we use the post-merger epoch of the last 5$\sigma$ data point from the detected lightcurve. The probability to detect kN decreases when increasing the required duration, which is expected as the optical luminosity monotonically decreases with time. The effects of the luminosity distance of the kN and the required observable time is illustrated in Fig.~\ref{fig:last5p}. We looked at the detection efficiency per limiting magnitude and observable window at 40, 100 and 200 Mpc. We find the minimum limiting magnitude required to ensure a 90\% probability of detection of a kN lasting 3 days is $<20.5$ at 40 Mpc, 21.5 at 100 Mpc and 23 at 200 Mpc. We estimate these limits averaging over the viewing angle distribution described in Section~\ref{subsec:transient_generator} and assuming the fiducial model $m_{\rm ej}=0.06\,\mathrm{M_{\sun}}$ and $\phi=60^{\circ}$.

\section{Feasibility for serendipitous detections}\label{sec:rates}

Next, we explore the feasibility for kN detections in time-domain surveys, irrespective of external GW triggers, for a volumetric BNS rate of 1000 Gpc$^{-3}$yr$^{-1}$, consistent with \cite{Scolnic2017}.
We make several simplifying assumptions, including losses due to candidate selection inefficiencies, bad weather, or other observing conditions. Every kNe observable within the covered volume with lightcurves of at least two 5$\sigma$ data points with a minimum separation of one hour is detected. Furthermore, a constant magnitude limit is assumed for the entire duration of an all-sky survey, i.e., 20\,000 sq.degrees. We assume that the distribution of BNS is homogeneous on the sky and rescale the number of detections using efficiencies calculated in Section \ref{sec:LVCtriggers}.
With the fiducial model, we find that for $m_{\rm lim}$ = 20.5 mag, 3 kNe are observable per year, around 20 for limiting magnitude of 22 mag  and 71 for 25 mag, assuming constant depth with $gri$ wavelength coverage (see Table \ref{tab:output_sim}).  Note that the optimistic assumptions make our results upper limits for the expected number of kNe detected for surveys with nominal depth from 20.5 to 25 mag observing with $gri$. 

\section{Summary and Conclusion}\label{sec:discussion}

The detection of GW-EM counterparts has proven to be very challenging. Kilonovae are both very rare and difficult to discover serendipitously, given their faintness and rapid brightness decline. The fast evolution in bolometric luminosity and color typically leads to lightcurves with few data points, hard to identify photometrically, and even more to get a good quality spectrum. In some cases, the lightcurve would fade below threshold within the first day from the merger. Fig. \ref{fig:last5p} shows the detection efficiency contours in the parameter space of limiting magnitude from 20.5 to 25 mag and required observable time window for the kN lightcurve from 0 to 7 days. For a given nominal depth, the probability of detecting kNe drops when requiring longer lightcurves. E.g., for $m_{\rm lim} = 23$ mag, there is more than 90\% probability for detecting a kilonova at 200 Mpc over 3 days since the merger. Requiring a duration of 5 days of the observable lightcurve, the efficiency drops down to about 10\% pushing to deeper observations of about 25 mag. For viewing angles towards the lanthanide-free component, reaching 22 mag in $gri$, and 23 mag for the viewing angles towards the lanthanide-rich component is sufficient to reach 90\% probability with a 3 days lightcurve.

In addition to the intrinsic observational challenges posed by the rapid evolution of kN lightcurves, the distances to the BNS low latency candidates triggered during the third run of LIGO and Virgo have been around 200 Mpc, about five times larger than for GW170817. Some of the events were detected with a single interferometer, which translates into poor localization. Large search areas, in combination with long distances, poses great challenges for the detection of EM counterparts. Survey scheduling under such conditions is even harder, considering the lack of knowledge of the underlying kN population. During O3, the effort to cover these large search areas on the sky was remarkable    {(e.g., \citealt{Antier2020a}; \citealt{Gompertz2020}; \citealt{Goldstein19}; \citealt{Kasliwal2020})}, but the adopted depth was generally insufficient to identify the possible counterpart at the reported distance, according to the models we have studied. Several collaborations such as DECam-GROWTH\footnote{\url{http://growth.caltech.edu/}} \citep{Anand2020}, ENGRAVE \citep{Ackley2020} or GRANDMA \citep{Antier2020b,Antier2020a} had triggered ToO observations with a global multi-telescope network when it was feasible. It has been shown that current capabilities can afford to rapidly cover large areas on the sky, which will lead to more constrainig observations with improved localizations in upcoming runs. 

Clearly, GW170817 appears to have been an exceptionally lucky occurrence, both with respect to distance, localisation and orientation, well outside the bulk of the BNS candidates in O3, as well as near the edge of the simulated sample  (see Fig.~\ref{fig:hist2d}). Due to their low rate, e.g., compared with other transients like supernovae, the chance of finding another nearby kN with viewing angle close to the pole, is very low. The asymmetry of the kN emission and the higher likelihood of viewing angles close to the merger plane mean that most of the kNe will be produced under conditions much less favourable for detection than GW170817.

In agreement with previous studies of follow up campaigns during O3 (\citealt{Coughlin2020}, \citealt{Hosseinzadeh2019}), we conclude that deeper observations are required to ensure  an efficient search for counterparts. At a distance of 200 Mpc, the required magnitude to observe a kN similar to AT2017gfo at lanthanide-rich viewing angles with detections over a period of three days is at least 23 mag. For an AT2017gfo-like event at that distance, pointing towards us with a viewing angle within the lanthanide-free component, $m_{\rm lim} = 22$ mag could be sufficient. This values change for different physical parameters of the kN model (see Section \ref{sec:lcs}).

The optical lightcurves of kNe become very red in just a few days. The nucleosynthesis of heavy nuclei translates into longer-lived emission at longer wavelengths. The blue emission fades rapidly, and it suffers more from absorption at viewing angles close to the merger plane, where the dynamical ejecta leaves most of the lanthanides. Thus, observations in red filters tend to be much more efficient. In this work, we show that adding $i$-band observations in an optical survey yields, on average, more than two times more detections than e.g., observing with only $g$- and $r$-band. The $g$-band lightcurves fade very quickly and, in some cases, would not even provide any detection. However, observations in bluer bands may play important roles in candidate vetting and characterization of kN events. Detection of the very fast decay in bluer bands, besides the longer-lasting redder wavelengths, is an excellent feature for the identification.

Ongoing ground-based telescopes can reach 23 mag under good atmospheric conditions and enough integration time on the source. However, due to the expected very fast decline of the luminosity, continued deep observations over multiple days is required, something that is hard to achieve with current instruments due to unpredictable weather patterns and moon phase among other factors. The Vera C. Rubin Observatory with the Legacy Survey of Space and Time \citep[LSST,][]{lsst} with its 8-meter-class telescope, planned to be online by 2023, will discover transients that are too far or too faint for the current available surveys, possibly including kNe \citep{Setzer19}.

We have used models tailored to BNS. However, we expect that results for NS-BH lightcurves would change quantitatively but not qualitatively.    Thus, we expect the suggested approach for optical follow-up should also apply to NS-BH LVC candidates. Our analyses show that the survey depths reached during O3 follow-ups were in general insufficient to guarantee high detection probability when accounting for the viewing angle dependence. Our results are independent on the nature of the LVC triggers and they can be used to scrutinize the observing strategies and to justify the non-detection for the candidates finally confirmed as BNS mergers.

In conclusion, detection and identification of GW electromagnetic counterparts from BNS mergers is very challenging. The survey magnitude depths during follow-up campaigns in O3 have generally been insufficient,  considering the asymmetry of the kN emission, the sometimes poor localization and distances, typically 5 times further than GW170817. Despite the lack of kN detections during O3, the multimessenger astronomy community have demonstrated a remarkable ability to respond promptly to LVC alerts, even those with wide search areas. Thus, the future is promising, also as more detectors join the gravitational waves observatory network, leading to better localization. Thus, we expect that with improved search strategies described here, future campaigns will be successful.

\section*{Acknowledgements}
The authors acknowledge support from the G.R.E.A.T research environment, funded by {\em Vetenskapsr\aa det},  the Swedish Research Council, project number 2016-06012.
The authors thank the referee for a thoughtful review that has significantly contributed to the readability and quality of the manuscript. 

\section*{Data availability}

The data underlying this article are derived from the Python package \texttt{simsurvey} found at \url{https://github.com/ZwickyTransientFacility/simsurvey}.

\texttt{possis}
The \textsc{possis} simulations are available at \url{https://github.com/mbulla/kilonova models}.

\bibliographystyle{mnras}
\bibliography{mnras}

\newpage
\appendix

\section{detection probability}\label{app:det_prob}

\begin{figure*}
    \centering
    \includegraphics[width=\textwidth]{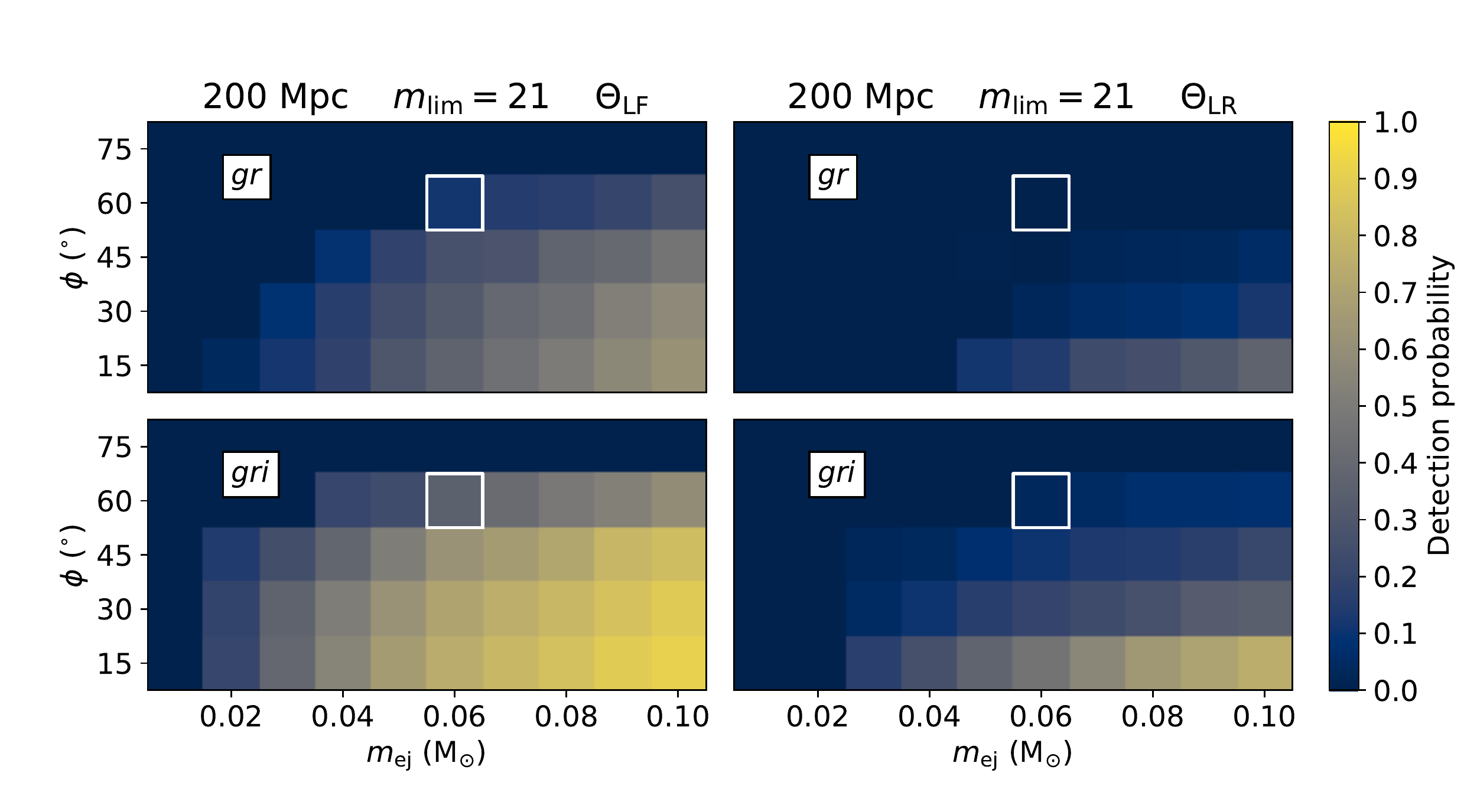}
    \includegraphics[width=\textwidth]{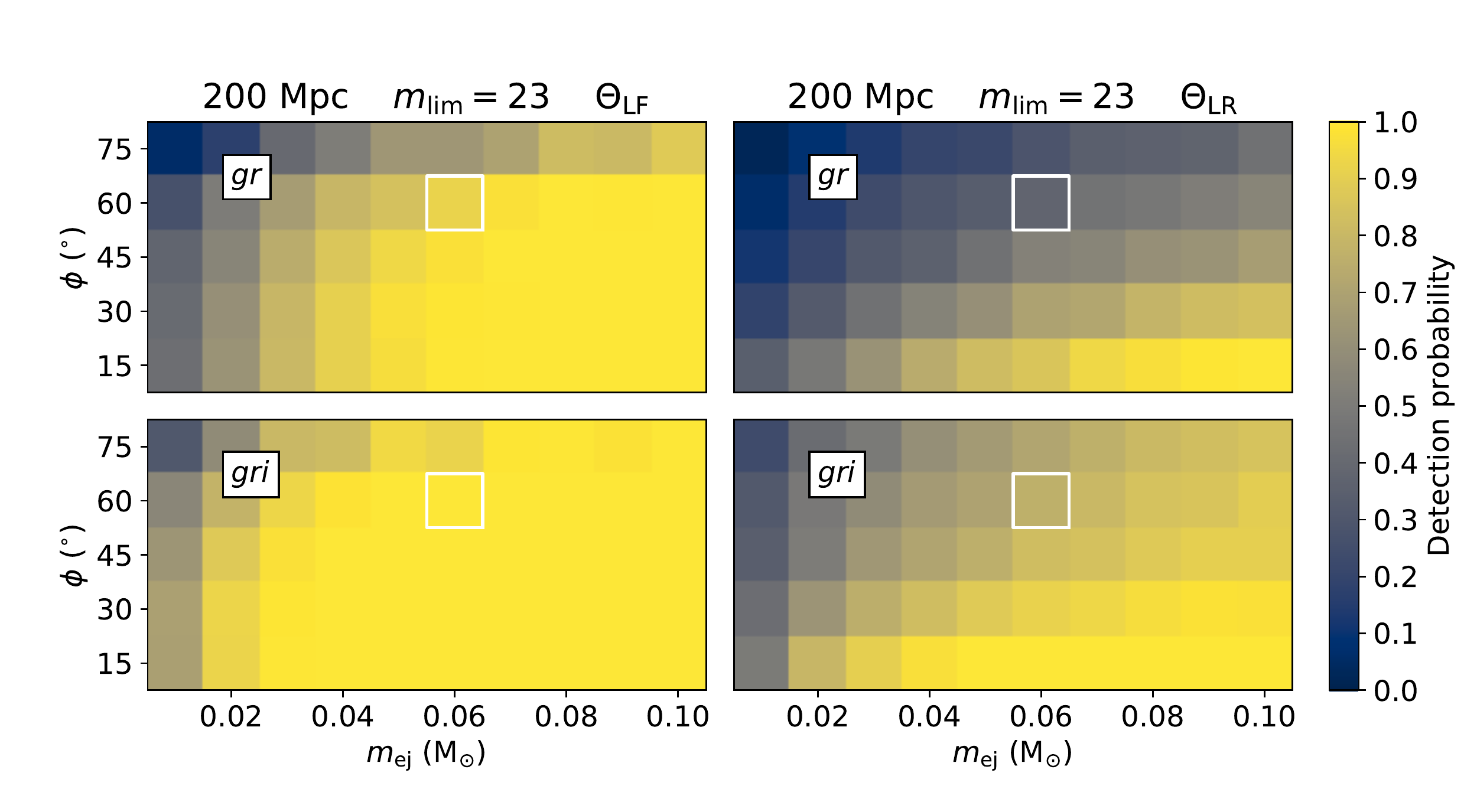}
    \caption{Detection probability for events at $\sim$200 Mpc. Probabilities are shown for observations with $gr$ and $gri$ with constant limiting magnitudes, 21 (top) and 23 (bottom) mag. Panel on the left side show probabilities for kN lightcurves observed with a viewing angle on the lanthanide-free region and lanthanide-rich on the right-hand side. The fiducial model is marked with a white rectangle.}
    \label{fig:p_21_23}
\end{figure*}

In section \ref{sec:all_grid}, we discuss how likely is a kilonova at 200 Mpc to be detected assuming constant depth in $gr$ and $gri$. We explore the entire parameter space of the models for ejecta masses ($m_{\rm ej}$) varying between 0.01 to 0.10 $\mathrm{M_{\sun}}$, half-opening angles ($\phi$) from 15 to 75 degrees and orientations towards the lanthanide-free ($\Theta_{\rm LF}$) or the lanthanide-rich ($\Theta_{\rm LR}$) components.
Fig. ~\ref{fig:p_22} presents the results for $m_{\rm lim} = 22$. Fig.~\ref{fig:p_21_23} shows results for $m_{\rm lim} = 21$ and $m_{\rm lim} = 23$ mag.

\bsp	
\label{lastpage}
\end{document}